\documentclass[aps, pra, onecolumn, amsmath, amssymb, superscriptaddress, nofootinbib]{revtex4-1}

\makeatletter
\def\p@subsection{}
\def\p@subsubsection{}
\makeatother

\usepackage[utf8]{inputenc}
\usepackage{graphicx}
\usepackage{units}
\usepackage{color}
\usepackage[pdftex, colorlinks=true, linkcolor=myblue, citecolor=myblue,
urlcolor=myblue]{hyperref}
\usepackage{times}
\usepackage{comment}
\usepackage{graphicx}
\usepackage{amsmath, calc}
\usepackage{mathrsfs}
\usepackage{amsfonts}
\usepackage{amssymb}
\usepackage{bm}
\usepackage{bbm}
\usepackage{color}
\usepackage{array}
\usepackage{units}
\usepackage{tabstackengine}
\stackMath

\definecolor{myblue}{rgb}{0,0,1}
\definecolor{myred}{rgb}{1,0,0}

\newcommand{\bra}[1]{\langle #1|}
\newcommand{\ket}[1]{|#1\rangle}

\newcommand{\expect}[1]{\langle #1 \rangle}

\DeclareMathOperator{\Tr}{Tr}

\begin{document}


\title{Parametrically driving a quantum oscillator into exceptionality}


\author{C. A. Downing}
\affiliation{Department of Physics and Astronomy, University of Exeter, Exeter EX4 4QL, United Kingdom}
\affiliation{E-mail: c.a.downing@exeter.ac.uk}

\author{A. Vidiella-Barranco}
\affiliation{Gleb Wataghin Institute of Physics, University of Campinas - UNICAMP, 13083-859, Campinas, SP, Brazil}


\date{\today}


\begin{abstract}
\noindent \textbf{Abstract}\\
The mathematical objects employed in physical theories do not always behave well. Einstein's theory of space and time allows for spacetime singularities and Van Hove singularities arise in condensed matter physics, while intensity, phase and polarization singularities pervade wave physics. Within dissipative systems governed by matrices, singularities occur at the exceptional points in parameter space whereby some eigenvalues and eigenvectors coalesce simultaneously. However, the nature of exceptional points arising in quantum systems described within an open quantum systems approach has been much less studied. Here we consider a quantum oscillator driven parametrically and subject to loss. This squeezed system exhibits an exceptional point in the dynamical equations describing its first and second moments, which acts as a borderland between two phases with distinctive physical consequences. In particular, we discuss how the populations, correlations, squeezed quadratures and optical spectra crucially depend on being above or below the exceptional point. We also remark upon the presence of a dissipative phase transition at a critical point, which is  associated with the closing of the Liouvillian gap. Our results invite the experimental probing of quantum resonators under two-photon driving, and perhaps a reappraisal of exceptional and critical points within dissipative quantum systems more generally.
\\
\\
\end{abstract}


\maketitle



\noindent \textbf{Introduction}\\
Conventionally, Hermitian physics has reigned supreme. Non-Hermitian extensions to standard theories were typically considered as mere perturbations, perhaps allowing for some small amount of dissipation to be accounted for. However, when a non-Hermitian matrix permits degeneracies in both its spectrum and its eigenfunctions at some special point -- a so-called exceptional point ($\mathrm{EP}$)~\cite{Kato1966} -- everything changes. Indeed, the physics arising in the vicinity of an $\mathrm{EP}$ may be significantly different to anything occuring in the Hermitian version of the theory~\cite{Berry2003, Heiss2003}. 

The basic ingredients underpinning $\mathrm{EP}$s ensures their widespread manifestation across large swathes of physics, including in acoustics~\cite{Shi2016, Fang2021}, mechanics~\cite{Mao2020, Rocha2020} and perhaps most famously in optics~\cite{Ozdemir2019, Miri2019}. More recently, $\mathrm{EP}$s arising in truly quantum mechanical systems have been studied using quantum master equations, which has revealed a certain richness beyond the realm of essentially classical $\mathrm{EP}$s~\cite{Chhajlany2019, Arkhipov2020, Nori2020, Avila2020, Arkhipov2021, Huybrechts2022}. In particular, the non-Hermitian quantum physics associated with parametrically driven systems has been shown to be rather captivating, especially for such formally simple systems, due to their close connection with parity-time symmetry and related concepts~\cite{Antonosyan2015, Ganainy2015, Miri2016, Wang2019, Zhang2021, Roy2021,You2021}.

Inspired by the recent studies of nonlinear, Kerr-like resonators with two-photon driving~\cite{Mirrahimi2014, Minganti2016, Bartolo2016, Savona2017, Puri2017, Lolli2017, Heugel2019, Nori2021, Mylnikov2022}, here we study a parametric driven-dissipative quantum oscillator. We place an emphasis on the squeezing and the phase-space representation of the quantum state of the oscillator, as well as both the first moments and the second moments of the system (including their steady state and transient behaviours). We link the $\mathrm{EP}$ arising in each case to a relevant observable quantity, including the mean populations~\cite{Zueco2021, Sturges2022}, optical spectrum, degrees of coherence and squeezed quadratures. We also highlight the importance of a critical point in the system, which is associated with a dynamical instability and a dissipative phase transition. The effects of anharmonicities on the oscillator are briefly discussed, mostly in relation to its effect on the closing of the Liouvillian gap and the presence of the dynamical instability, since a full treatment of such a nonlinearity can be found in Refs.~\cite{Mirrahimi2014, Minganti2016, Bartolo2016, Savona2017, Puri2017, Lolli2017, Heugel2019, Nori2021, Mylnikov2022}.

The studied driven-dissipative parametric system may be realized in modern quantum optical laboratories with judicious use of cavities and traps~\cite{Leghtas2015, Wang2016, Ding2017, Pechal2019}, or by designing certain types of nonlinear quantum circuits~\cite{Wustmann2019, Liu2020, Chien2020}. The discovery of $\mathrm{EP}$s in such a parametric system would join a growing list of quantum systems, mostly involving superconducting qubits but also including trapped-ion systems, which exhibit exceptional point physics~\cite{Naghiloo2019, Partanen2019, Dogra2021, Chen2021, Chen2022, Liang2023, Li2023, Quinn2023}. Already, $\mathrm{EP}$s have been experimentally shown to be useful for the control of quantum states and even for the enhancement of quantum heat engines~\cite{Abbasi2022, Zhang2022, Bu2023}.

\begin{figure*}[tb]
 \includegraphics[width=\linewidth]{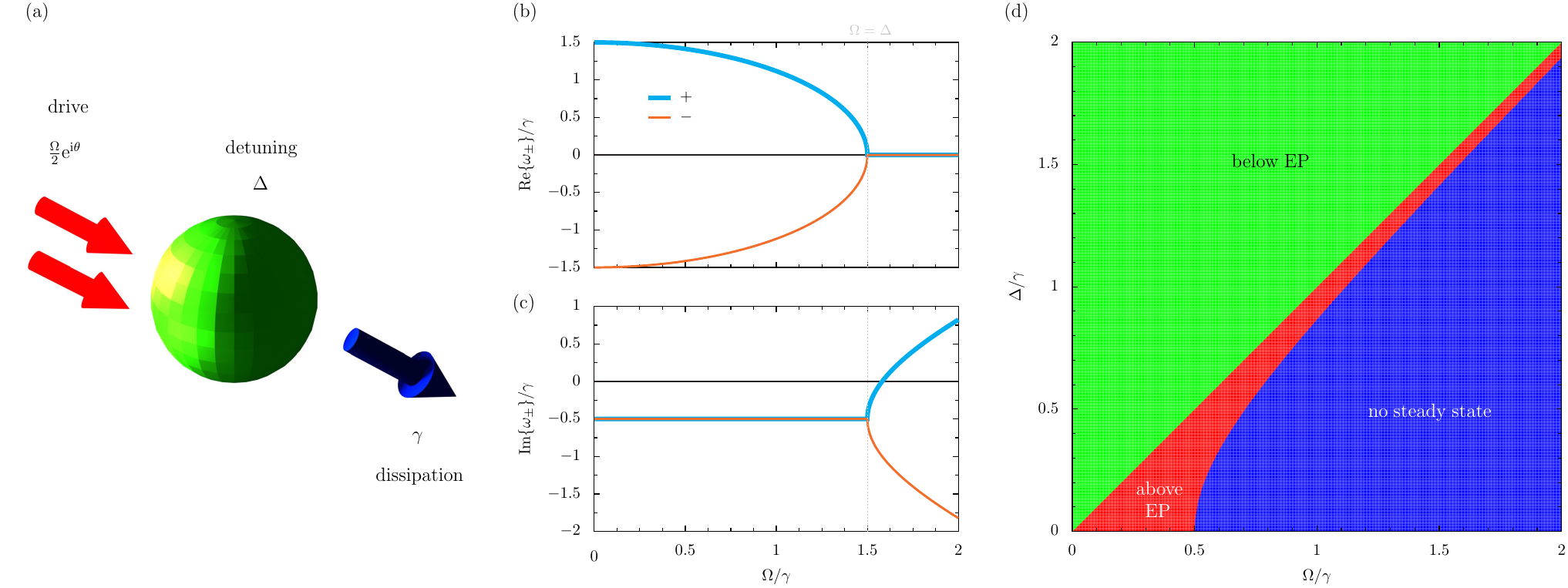}
 \caption{ \textbf{Fundamentals of the parametric driven-dissipative oscillator.} Panel (a): a sketch of the quantum oscillator (green ball), being driven coherently (two red arrows) with an amplitude $\Omega$ and phase $\theta$, and where the detuning is $\Delta$ [cf. Eq.~\eqref{eq:Hammy}]. The oscillator suffers losses (blue arrow) at the rate $\gamma$ [cf. Eq.~\eqref{eqapp:massdsdsdter}]. Panel (b): the real parts of the eigenvalues $\omega_{\pm}$, arising from the first moments matrix (or effective Hamiltonian) $\mathcal{H}$, as a function of the drive amplitude $\Omega$ (in units of $\gamma$) [cf. Eq.~\eqref{eqapp:umadfdfsdsdsdsdstrix}]. Panel (c): the corresponding imaginary parts of $\omega_{\pm}$. Dashed grey lines: the exceptional point (where $\Omega = \Delta$). Panel (d): the phase diagram of the system as a function of $\Delta$ and $\Omega$, where the green region is below the exceptional point ($\Omega < \Delta $) and the red region is above the exceptional point ($\Omega >  \Delta $) [cf. Eq.~\eqref{eqapp:fghn}]. No steady state forms in the blue region [cf. Eq.~\eqref{eqapp:dfgdsfdfx}]. In this figure, we consider the case of $\Delta = 3\gamma/2$, so that the exceptional driving amplitude $\Omega_{\mathrm{EP}} = 3\gamma /2$ [cf. Eq.~\eqref{eqapp:fghn}] and the critical driving amplitude $\Omega_{\mathrm{c}} = \sqrt{5/2}\gamma \simeq 1.58 \gamma$ [cf. Eq.~\eqref{eqapp:dfgdssdsfdfx}].}
 \label{excep}
\end{figure*}

The parametric driven-dissipative quantum harmonic oscillator model, as sketched in Fig.~\ref{excep}~(a), may be described by the following Hamiltonian $\hat{H}$ (here and throughout we take $\hbar = 1$)~\cite{Loudon1987}
\begin{equation}
\label{eq:Hammy}
 \hat{H} = \Delta b^{\dagger} b 
 + \frac{\Omega}{2} \mathrm{e}^{\mathrm{i} \theta} b^{\dagger} b^{\dagger} +  \frac{\Omega}{2} \mathrm{e}^{-\mathrm{i} \theta} b b,
\end{equation}
where, without loss of generality, the oscillator-driving detuning $\Delta \ge 0$, the two-excitation driving amplitude $\Omega  \ge 0$, and the driving phase $-\pi \le \theta \le \pi$ (see the Supplementary Information for more details). The creation and annihilation ladder operators $b^\dagger$ and $b$ act on a number state $\ket{n}$ as follows: $b^\dagger \ket{n} = \sqrt{n+1} \ket{n}$ and $b \ket{n} = \sqrt{n} \ket{n-1}$, and they obey the bosonic commutation relation $[ b, b^\dagger] = 1$. The diagonalization of the Hamiltonian $\hat{H}$ of Eq.~\eqref{eq:Hammy} is possible with the aid of the bosonic Bogoliubov operator $\beta$, defined in terms of the squeezing parameter $\phi$ as
\begin{equation}
\label{eq:Hamm213sdfsdf131232y}
\beta = \cosh \left( \phi \right) b + \mathrm{e}^{\mathrm{i} \theta} \sinh \left( \phi \right) b^\dagger,
\quad\quad\quad\quad\quad\quad \quad
\phi = \frac{\ln \left( \frac{\Delta+\Omega}{\Delta-\Omega} \right)}{4},
\end{equation}
which satisfies the commutator $[ \beta, \beta^\dagger] = 1$, and which is valid for sufficiently small driving amplitudes such that $\Omega < \Delta$~\cite{Tsallis1978, Colpa1978}. This Bogoliubov transformation leads to the diagonalized form of the Hamiltonian $\hat{H}$, complete with Bogoliubov mode eigenfrequency $\tilde{\omega}$, as follows
\begin{equation}
\label{eq:Hamm213131232y}
 \hat{H} = \tilde{\omega} \beta^{\dagger} \beta, 
\quad\quad\quad\quad\quad\quad\quad\quad
 \tilde{\omega} = \sqrt{\Delta^2-\Omega^2}.
\end{equation}
This brief analysis reveals a squeezed energy ladder with the eigenfrequencies $E_n = n \tilde{\omega}$, which correspond to the squeezed number states $\ket{n, \phi} = \mathcal{S}_\phi \ket{n}$, where the squeezing operator $\mathcal{S}_\phi = \exp{\left( \frac{1}{2} \phi \mathrm{e}^{-\mathrm{i} \theta} b b  -\frac{1}{2} \phi \mathrm{e}^{\mathrm{i} \theta} b^{\dagger} b^{\dagger}  \right)}$. The squeezed oscillator levels $E_n$ are separated by the spacing $\Delta$ at small driving amplitudes (namely, the harmonic oscillator limit with $\Omega \ll \Delta$), while the inter-level spacing is vanishing in the limit of large driving amplitudes ($\Omega \to \Delta$). Notably, there is so-called ``spectral collapse'' for $\Omega > \Delta$, where the eigenfrequencies $E_n$ become complex [in keeping with the construction of the squeezing parameter $\phi$ in Eq.~\eqref{eq:Hamm213sdfsdf131232y}], however we do not enter this regime within the closed system version of our theory as solely described by Eq.~\eqref{eq:Hammy}.

Interestingly, the quantum oscillator population $\langle n, \phi | b^\dagger b | n, \phi \rangle = n \cosh \left( 2 \phi \right) + \sinh^2 \left( \phi \right)$, so that the Bogoliubov ground state $\ket{0, \phi}$ has a nonzero population $\langle 0, \phi | b^\dagger b | 0, \phi \rangle = \sinh^2 \left( \phi \right)$ due to the parametric driving. With the twin definitions of dimensionless position $\hat{X} = ( \mathrm{e}^{\mathrm{i}\theta/2} b^\dagger + \mathrm{e}^{-\mathrm{i}\theta/2} b )/\sqrt{2}$ and dimensionless momentum $\hat{P} = \mathrm{i} ( \mathrm{e}^{\mathrm{i}\theta/2} b^\dagger - \mathrm{e}^{-\mathrm{i}\theta/2} b )/\sqrt{2}$, the normalization ensures that the quadrature commutation relation $[ \hat{X}, \hat{P} ] = \mathrm{i}$ holds, while we have chosen the phase $\theta/2$ in these definitions in order to compensate for the phase $\theta$ appearing the drive term [cf. Eq.~\eqref{eq:Hammy}]. The mean value of the complex field amplitude in the vacuum squeezed state is zero, so the quadrature expectation values $\langle 0, \phi | \hat{X} | 0, \phi \rangle = \langle 0, \phi | \hat{P} | 0, \phi \rangle = 0$. However, the squared counterparts $\langle 0, \phi | \hat{X}^2 | 0, \phi \rangle = \mathrm{e}^{-2\phi}/2$ and $\langle 0, \phi | \hat{P}^2 | 0, \phi \rangle = \mathrm{e}^{2\phi}/2$, highlighting the quadrature squeezing as governed by the squeezing parameter $\phi$ [cf. Eq.~\eqref{eq:Hamm213sdfsdf131232y}]. These closed system results follow from the inverse relation of Eq.~\eqref{eq:Hamm213sdfsdf131232y}, where $b = \cosh \left( \phi \right) \beta - \mathrm{e}^{\mathrm{i} \theta} \sinh \left( \phi \right) \beta^\dagger$.

We include dissipation in the model via an open quantum systems approach. Employing the Born, Markov and secular approximations, the associated quantum master equation for the system's reduced density matrix $\rho$ may be given by~\cite{Breuer2002, Gardiner2014}
\begin{equation}
\label{eqapp:massdsdsdter}
 \partial_t \rho = \mathrm{i} [ \rho, \hat{H} ]
+  \frac{\gamma}{2} \left( 2 b \rho b^{\dagger} -  b^{\dagger} b \rho - \rho b^{\dagger} b \right),
\end{equation}
where $\gamma \ge 0$ is the dissipation rate and the Hamiltonian $\hat{H}$ is given by Eq.~\eqref{eq:Hammy}. This quantum model includes the unitary evolution of the density matrix $\rho$ via the the first term on the right-hand-side of Eq.~\eqref{eqapp:massdsdsdter} -- the Liouville–von Neumann equation -- while non-unitary evolution is accounted for by the second term in Eq.~\eqref{eqapp:massdsdsdter}, which upgrades Eq.~\eqref{eqapp:massdsdsdter} to the form of a Gorini–Kossakowski–Sudarshan–Lindblad equation. We take $\gamma$ as a phenomenological parameter, but in the case of modelling a thermal bath $\gamma$ is the zero-temperature decay rate, and we show in the Supplementary Information that the effects of nonzero temperatures are rather negligible (for example, the location of the $\mathrm{EP}$ is unchanged).

Using the property $\Tr{ \left( \mathcal{O} \rho \right) } = \expect{\mathcal{O}}$, which leads to the mean value $\expect{\mathcal{O}}$ of any operator $\mathcal{O}$, the averages of the first moments of the system (like $\langle  b  \rangle$ for example) may be readily found. For the presented parametric model, as defined by Eq.~\eqref{eqapp:massdsdsdter} supplemented with Eq.~\eqref{eq:Hammy}, this procedure leads to the following Schr\"{o}dinger-like dynamical equation
\begin{equation}
\label{eqapp:umasdsdsdsdstrix}
\mathrm{i} \partial_t \psi = \mathcal{H} \psi,
 \end{equation}
 where $\psi$, the two-dimensional column vector which collects the first moments, and the dynamical matrix $\mathcal{H}$ are defined via
 \begin{equation}
\label{eqapp:umasdsdsdsdstrix56456}
\psi = 
\begin{pmatrix}
\langle b \rangle  \\
\langle b^\dagger \rangle
\end{pmatrix},
\quad\quad\quad\quad\quad\quad
\mathcal{H} =
\begin{pmatrix}
\Delta - \mathrm{i} \frac{\gamma}{2} & \Omega \mathrm{e}^{\mathrm{i} \theta}  \\
-\Omega \mathrm{e}^{-\mathrm{i} \theta} & -\Delta - \mathrm{i} \frac{\gamma}{2}
\end{pmatrix}. 
 \end{equation}
 The two complex eigenfrequencies $\omega_{\pm}$ arising from the $2 \times 2$ matrix effective Hamiltonian $\mathcal{H}$ read
  \begin{equation}
\label{eqapp:umadfdfsdsdsdsdstrix}
\omega_{\pm} = \begin{cases}
- \mathrm{i} \frac{\gamma}{2} \pm \tilde{\omega},  & \quad\quad \left( \Omega < \Delta \right), \\
- \mathrm{i} \frac{\gamma}{2},  & \quad\quad \left( \Omega = \Delta \right), \\
- \mathrm{i} \left( \frac{\gamma}{2} \pm \Gamma \right), & \quad\quad \left( \Omega > \Delta \right),
\end{cases}
 \end{equation}
which generalizes the closed Hamiltonian $\hat{H}$ description of Eq.~\eqref{eq:Hamm213131232y}, with its wholly real energies $ \tilde{\omega}$ and its restriction $\Omega < \Delta$. In particular, we have introduced the effective decay rate $\Gamma$, defined via the relation [cf. the definition of the frequency $ \tilde{\omega}$ from Eq.~\eqref{eq:Hamm213131232y}]
\begin{equation}
\label{eqapp:dfcbfcfbddsfdsfdfx}
 \Gamma = \sqrt{\Omega^2-\Delta^2},
\end{equation}
which captures the physics of system in its dissipative phase with purely imaginary eigenvalues $\omega_{\pm}$. The two normalized eigenvectors $\alpha_{\pm}$, corresponding to the eigenfrequencies $\omega_{\pm}$, are given by
   \begin{equation}
\label{eqapp:fdfd}
\alpha_{\pm} = \frac{1}{\sqrt{ 2}}
\begin{pmatrix}
 \frac{\Delta \pm \tilde{\omega}}{\Omega}    \\
-  \mathrm{e}^{-\mathrm{i} \theta}
\end{pmatrix}, 
\quad\quad\quad\quad
\left( \Omega < \Delta \right),
\quad\quad\quad\quad
\alpha_{\pm} = \frac{1}{\sqrt{ 2} }
\begin{pmatrix}
 \frac{\Delta \mp \mathrm{i} \Gamma}{\Omega}    \\
-  \mathrm{e}^{-\mathrm{i} \theta}
\end{pmatrix}, 
\quad\quad\quad\quad
\left( \Omega > \Delta \right).
 \end{equation} 
Clearly, both Eq.~\eqref{eqapp:umadfdfsdsdsdsdstrix} and Eq.~\eqref{eqapp:fdfd} present a point of coalescence -- a so-called $\mathrm{EP}$ -- at the particular driving amplitude $\Omega = \Omega_{\mathrm{EP}}$, defined by
  \begin{equation}
\label{eqapp:fghn}
  \Omega_{\mathrm{EP}} =  \Delta .
 \end{equation}
At this special point the eigenfrequency $\tilde{\omega} = 0$ [cf. Eq.~\eqref{eq:Hamm213131232y}], and the effective Hamiltonian $\mathcal{H}$ in Eq.~\eqref{eqapp:umasdsdsdsdstrix} becomes defective~\cite{Berry2003, Heiss2003}. Since the $\mathrm{EP}$ of Eq.~\eqref{eqapp:fghn} has arisen from the Hamiltonian-like analysis of the dynamical equation in Eq.~\eqref{eqapp:umasdsdsdsdstrix}, it can be considered to be a so-called `Hamiltonian $\mathrm{EP}$'. Using the eigenfrequencies of Eq.~\eqref{eqapp:umadfdfsdsdsdsdstrix}, we plot in Fig.~\ref{excep}~(b, c) the real and imaginary parts of $\omega_{\pm}$ as a function of the drive amplitude $\Omega$ (for an example case with the detuning $\Delta = 3\gamma/2$). The $\mathrm{EP}$ is marked by the vertical, dashed grey line in both panels and highlights the bifurcations in the complex eigenvalues $\omega_{\pm}$, as well as the guarding of the boundary between two drastically different physical phases. In what follows, we shall consider the consequences of this $\mathrm{EP}$ for the physical responses of the squeezed oscillator system.
\\

\begin{figure*}[tb]
 \includegraphics[width=\linewidth]{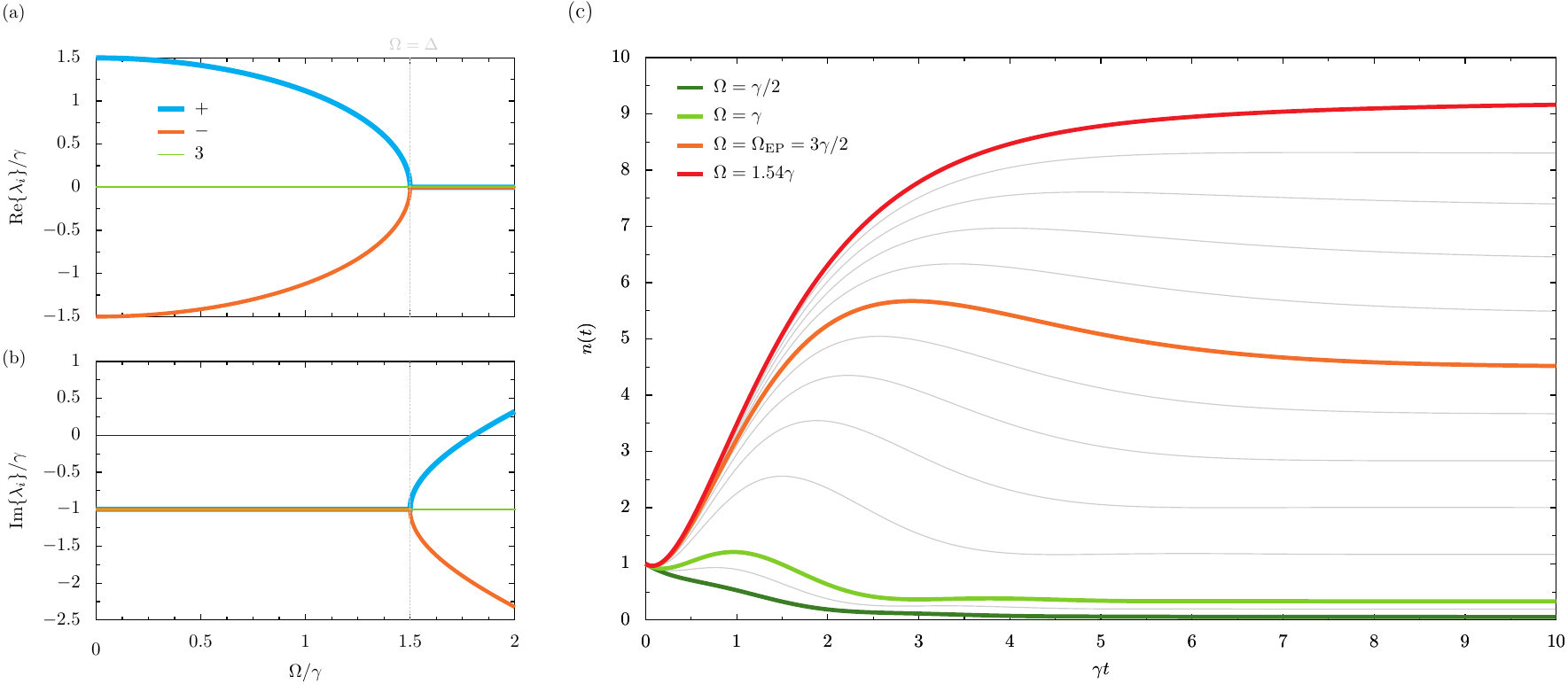}
 \caption{ \textbf{Populations of the parametric driven-dissipative oscillator.} Panel (a): the real parts of the eigenvalues $\lambda_{i}$, arising from the second moments matrix $\mathcal{M}$, as a function of the drive amplitude $\Omega$ (in units of the loss rate $\gamma$) [cf. Eq.~\eqref{eqapp:sdffaf}]. Panel (b): the corresponding imaginary parts of $\lambda_{i}$. Dashed grey lines: the exceptional point (where $\Omega = \Delta$). Panel (c): the mean population of the oscillator $n(t)$, as a function of time $t$ (in units of $\gamma^{-1}$) with the initial mean population $n(0) = 1$ due to the oscillator being in its first excited state [cf. Eq.~\eqref{eqapp:dfgddsfdsfdfx}]. We consider several values of $\Omega$, including below the exceptional point (green lines) and above the exceptional point (red line). The result exactly at the exceptional point is given by the orange line [cf. Eq.~\eqref{eqapp:adsads}]. Thin grey lines: intermediate values of $\Omega$. In this figure, we consider the case of the detuning $\Delta = 3\gamma/2$, so that $\Omega_{\mathrm{EP}} = 3\gamma /2$ [cf. Eq.~\eqref{eqapp:fghn}] and $\Omega_{\mathrm{c}} = \sqrt{5/2}\gamma \simeq 1.58 \gamma$ [cf. Eq.~\eqref{eqapp:dfgdssdsfdfx}].}
 \label{poppy}
\end{figure*}

\noindent \textbf{Results}\\
Similar to the calculation leading to the coupled dynamical equations of Eq.~\eqref{eqapp:umasdsdsdsdstrix}, the second moments of the system (that is quantities like $\langle  b^\dagger b  \rangle$, the average population of the oscillator) are defined through the non-homogeneous equation
\begin{equation}
\label{eqapp:dfdfdfsfdav}
\mathrm{i} \partial_t \Psi = \mathcal{M} \Psi + \mathcal{P}, 
 \end{equation}
where the three-dimensional column vector $\Psi$, which gathers the second moments, the driving term $\mathcal{P}$ and the $3 \times 3$ matrix $\mathcal{M}$ read [cf. Eq.~\eqref{eqapp:umasdsdsdsdstrix56456}]
\begin{equation}
\label{eqapp:zzzscsd}
\Psi = 
\begin{pmatrix}
\langle  b^\dagger b  \rangle   \\
\langle  b b  \rangle \\
\langle  b^\dagger b^\dagger
\end{pmatrix},
\quad\quad\quad\quad
\mathcal{M} =
\begin{pmatrix}
 -\mathrm{i} \gamma & -\Omega  \mathrm{e}^{-\mathrm{i} \theta} &  \Omega  \mathrm{e}^{\mathrm{i} \theta}  \\
 2  \Omega  \mathrm{e}^{\mathrm{i} \theta} &  2  \Delta  -\mathrm{i} \gamma &   0  \\
  -2 \Omega  \mathrm{e}^{-\mathrm{i} \theta} & 0 &  - 2  \Delta -\mathrm{i} \gamma
 \end{pmatrix},
 \quad\quad\quad\quad
\mathcal{P} =
\begin{pmatrix}
0   \\
 \Omega \mathrm{e}^{\mathrm{i} \theta} \\
- \Omega \mathrm{e}^{-\mathrm{i} \theta}
\end{pmatrix}.
 \end{equation}
 The three complex eigenvalues $\lambda_{+}, \lambda_{-}$ and $\lambda_{3}$ of the dynamical matrix $\mathcal{M}$ are given by [cf. Eq.~\eqref{eqapp:umadfdfsdsdsdsdstrix}]
  \begin{equation}
\label{eqapp:sdffaf}
\lambda_{3} = - \mathrm{i} \gamma,
\quad\quad\quad\quad\quad\quad\quad
\lambda_{\pm} = \begin{cases}
 - \mathrm{i} \gamma \pm \tilde{\omega},  & \quad\quad \left( \Omega < \Omega_{\mathrm{EP}} \right), \\
 - \mathrm{i} \gamma,  & \quad\quad \left( \Omega = \Omega_{\mathrm{EP}} \right), \\
 - \mathrm{i} \left( \gamma \pm \Gamma \right), & \quad\quad \left( \Omega > \Omega_{\mathrm{EP}} \right),
\end{cases}
 \end{equation}
where $\tilde{\omega}$ and $\Gamma$ are defined in Eq.~\eqref{eq:Hamm213131232y} and Eq.~\eqref{eqapp:dfcbfcfbddsfdsfdfx} respectively. The three corresponding (and unnormalized) eigenvectors $\beta_{3}, \beta_{+}$ and $\beta_{-}$ read (we assume $\Omega < \Omega_{\mathrm{EP}} $ for the moment, the results for $\Omega > \Omega_{\mathrm{EP}} $ are of course similar)
   \begin{equation}
\label{eqapp:sdffxfdsdaf}
\beta_{3} = 
\begin{pmatrix}
-\Delta   \\
\Omega  \mathrm{e}^{\mathrm{i} \theta} \\
~~\Omega  \mathrm{e}^{-\mathrm{i} \theta}
\end{pmatrix},
\quad\quad\quad\quad\quad\quad\quad
\beta_{\pm} = 
\begin{pmatrix}
 -\Omega \left( \Delta \pm \tilde{\omega} \right)   \\
\left[ 2\Delta \left( \Delta \pm \tilde{\omega} \right) - \Omega^2 \right] \mathrm{e}^{\mathrm{i} \theta} \\
\Omega^2 \mathrm{e}^{-\mathrm{i} \theta}
\end{pmatrix}.
 \end{equation}
The simultaneous coalescing of all three eigenvalues $\lambda_i$ and eigenvectors $\beta_i$ at the certain point $\Omega = \Omega_{\mathrm{EP}}$ reveals that the $\mathrm{EP}$ in the second moments occurs at exactly the same point in parameter space as the $\mathrm{EP}$ in the first moments  [cf. Eq.~\eqref{eqapp:sdffaf} and Eq.~\eqref{eqapp:sdffxfdsdaf} with Eq.~\eqref{eqapp:fghn}]. This suggests that the `Hamiltonian $\mathrm{EP}$' [arising from Eq.~\eqref{eqapp:umasdsdsdsdstrix}] and `Liouvillian $\mathrm{EP}$' [arising from Eq.~\eqref{eqapp:dfdfdfsfdav}] are identical for this simple system. Furthermore, the second moments $\mathrm{EP}$ may be classed as being of third-order (due to its emergence with the coalescing of three objects), while the first moments are associated with a second-order $\mathrm{EP}$ (only two objects coalesce in this case)~\cite{Hodaei2017, Downing2021}. The higher-order $\mathrm{EP}$ associated with Eq.~\eqref{eqapp:dfdfdfsfdav} is graphed in Fig.~\ref{poppy}~(a, b) using the results of Eq.~\eqref{eqapp:sdffaf}, where the real parts of the eigenvalues $\lambda_i$ are displayed in the upper panel, and the corresponding imaginary parts in the lower panel. This analysis shows the importance of exceptional point physics throughout the different levels of description of open quantum systems (for example with increasing large moments).
\\

\noindent \textit{Average populations}.\\
 The mean population of the oscillator $n(t) = \langle b^\dagger b  \rangle$ is contained within the first element of $\Psi$ [cf. Eq.~\eqref{eqapp:zzzscsd}] and is found by formally solving the equation of motion defined in Eq.~\eqref{eqapp:dfdfdfsfdav}. At long time scales ($t \to \infty$), the steady state population $n(\infty) = \lim_{t \to \infty} \langle b^\dagger b  \rangle$ readily follows from Eq.~\eqref{eqapp:dfdfdfsfdav} by noting that in this limit $\partial_t \Psi = 0$ and so its steady state solution $- \mathcal{M}^{-1} \mathcal{P}$ yields
\begin{equation}
\label{eqapp:dfgdsfdfx}
n(\infty) = \frac{1}{2} \frac{\Omega^2}{   \tilde{\omega}^2  + \left( \frac{\gamma}{2} \right)^2} = \frac{1}{2} \frac{\Omega^2}{ \left( \frac{\gamma}{2} \right)^2 - \Gamma^2},
\end{equation}
where $\tilde{\omega}$ is defined in Eq.~\eqref{eq:Hamm213131232y} and $\Gamma$ in Eq.~\eqref{eqapp:dfcbfcfbddsfdsfdfx}. As a population, the denominator of Eq.~\eqref{eqapp:dfgdsfdfx} should be non-negative. This suggests that the presented system is stable for driving amplitudes $\Omega < \Omega_{\mathrm{c}}$, where the critical driving frequency $\Omega_{\mathrm{c}}$ is defined as
\begin{equation}
\label{eqapp:dfgdssdsfdfx}
\Omega_{\mathrm{c}} = \sqrt{\Delta^2  + \left( \frac{\gamma}{2} \right)^2}.
\end{equation}
Otherwise there is no steady state -- the competition from the driving and the dissipation is conclusively won by the driving and the population increases in time without bound. In practice, this continual climbing of the infinite and bosonic energy ladder of the oscillator can be tamed by either truncating the oscillator or by considering anharmoncities (both cases are discussed later on, although we are not so concerned with driving amplitudes satisfying $\Omega \ge \Omega_{\mathrm{c}}$ since the $\mathrm{EP}$ has already been passed by this stage). Notably, the result of Eq.~\eqref{eqapp:dfgdssdsfdfx} was foreshadowed by the complex eigenvalues $\omega_{\pm}$ provided in Eq.~\eqref{eqapp:umadfdfsdsdsdsdstrix}. In particular, in the regime of $\Omega > \Delta$ one of the eigensolutions ($\omega_{-}$) corresponds to exponentially growing behaviour in time when $\gamma/2 < \Gamma$, consistent with the relation of Eq.~\eqref{eqapp:dfgdssdsfdfx}.

The steady state phase diagram of the system implied by Eq.~\eqref{eqapp:dfgdssdsfdfx}, as a function of the detuning $ \Delta $ and the driving amplitude $\Omega$, is shown in Fig.~\ref{excep}~(d). The green region occurs below the $\mathrm{EP}$ ($\Omega <  \Delta$), while the red region arises above the $\mathrm{EP}$ ($\Omega >  \Delta $). No steady state is able to form in the blue region, as suggested by Eq.~\eqref{eqapp:dfgdssdsfdfx}, and hence the population here is unbounded with increasing time. Hence for vanishing detuning $ \Delta  \ll \gamma$ the steady state most likely exists in the `above $\mathrm{EP}$' red region, within the bound $0 < \Omega < \gamma/2$. Contrarily, with large detuning $ \Delta  \gg \gamma$ the steady state is most likely to be in the `below $\mathrm{EP}$' green region $ \Delta  > \Omega$, with a smaller chance of being in the slither of `above $\mathrm{EP}$' red region when $ \Delta  < \Omega <  \Delta  \left( 1 + \gamma^2/8\Delta^2 \right)$. The diagram of Fig.~\ref{excep}~(d) thus acts as a kind of map for the parametric oscillator, highlighting both the exceptional point at $\Omega = \Omega_{\mathrm{EP}}$ and the critical point at $\Omega = \Omega_{\mathrm{c}}$.

The full solution of the equation of motion given in Eq.~\eqref{eqapp:dfdfdfsfdav} leads to the oscillator population $n (t)$, including both transient and steady state parts, which in general is given by
\begin{equation}
\label{eqapp:dfgddsfdsfdfx}
n (t) = n(\infty) + \frac{\tilde{\omega}^{-2}/4}{ \tilde{\omega}^2 + \left(\tfrac{\gamma}{2}\right)^2} \bigg\{ 4 n(0) \Delta^2 \Big[ \tilde{\omega}^2 + \left(\tfrac{\gamma}{2}\right)^2 \Big] - \gamma \tilde{\omega} \Omega^2 \sin \left( 2 \tilde{\omega} t \right) - 2 \Omega^2 \bigg(  2 n(0) \Big[ \tilde{\omega}^2 + \left(\tfrac{\gamma}{2}\right)^2 \Big] + \tilde{\omega}^2 \bigg) \cos \left( 2 \tilde{\omega} t \right) \bigg\}  \mathrm{e}^{-\gamma t}, 
\end{equation}
for driving amplitudes $\Omega < \Omega_{\mathrm{EP}}$, where the frequency $\tilde{\omega}$ is defined in Eq.~\eqref{eq:Hamm213131232y}, and where $n(0)$ is the population of the oscillator at the initial time $t = 0$. Similarly, for driving amplitudes $\Omega > \Omega_{\mathrm{EP}}$ the analogous result, using the damping rate $\Gamma$ as given in Eq.~\eqref{eqapp:dfcbfcfbddsfdsfdfx}, is
\begin{equation}
\label{eqapp:dfgsdfsfddsfdsfdfx}
n (t) = n(\infty) - \frac{\Gamma^{-2}/4}{ \left(\tfrac{\gamma}{2}\right)^2 -\Gamma^2} \bigg\{ 4 n(0) \Delta^2 \Big[  \left(\tfrac{\gamma}{2}\right)^2 - \Gamma^2 \Big] + \gamma \Gamma \Omega^2 \sinh \left( 2 \Gamma t \right) - 2 \Omega^2 \bigg( 2 n(0) \Big[  \left(\tfrac{\gamma}{2}\right)^2 - \Gamma^2 \Big] - \Gamma^2 \bigg) \cosh \left( 2 \Gamma t \right) \bigg\}  \mathrm{e}^{-\gamma t}. 
\end{equation}
The solution of Eq.~\eqref{eqapp:dfgddsfdsfdfx} displays characteristic sinusoidal and cosinusoidal Rabi-like oscillations, along with an exponential decay with the time constant $\gamma$, until the driving amplitude $\Omega$ overcomes the detuning $\Delta$ and the nonoscillatory solution of Eq.~\eqref{eqapp:dfgsdfsfddsfdsfdfx} supersedes it. However, when the driving amplitude is exactly $\Omega = \Omega_{\mathrm{EP}}$ [cf. Eq.~\eqref{eqapp:fghn}] the solution of Eq.~\eqref{eqapp:dfgddsfdsfdfx} is drastically reconstructed into the much simpler form
\begin{equation}
\label{eqapp:adsads}
n_{\mathrm{EP}} (t) = \frac{2\Delta^2}{\gamma^2} + \bigg\{ n(0) + \frac{2\Delta^2}{\gamma^2} \bigg( \gamma t \Big[ n(0) \gamma t - 1 \Big] -1 \bigg) \bigg\}  \mathrm{e}^{-\gamma t}, \quad\quad\quad\quad\quad\quad
 \left( \Omega = \Omega_{\mathrm{EP}} \right).
\end{equation}
Most notably, while the exponential decay is unaffected, the expression of Eq.~\eqref{eqapp:adsads} features both linear and quadratic terms in the dimensionless time $\gamma t$ (instead of this quantity only appearing in trigonometric or hyperbolic functions) due to the nature of the $\mathrm{EP}$. We plot the average population $n(t)$ in Fig.~\ref{poppy}~(c) for several values of driving amplitude $\Omega$, where the detuning is fixed at $\Delta = 3\gamma/2$ (so that $\Omega_{\mathrm{EP}} = 3\gamma /2$ and $\Omega_{\mathrm{c}} = \sqrt{5/2}\gamma \simeq 1.58 \gamma$ from Eq.~\eqref{eqapp:fghn} and Eq.~\eqref{eqapp:dfgdssdsfdfx} respectively). For small driving amplitudes $\Omega$ -- well below the $\mathrm{EP}$ (dark green line) -- the population of the oscillator never becomes significant and monotonically decreases to a low plateau. With increasing drive amplitudes $\Omega$ a non-monotonic behaviour develops in the manner of Rabi oscillations (lime green line), which is more pronounced for larger drives, for example approaching the $\mathrm{EP}$ (orange line). Above the $\mathrm{EP}$ (red line) the population surges to increasingly large steady state plateaus without oscillation. Finally, above the critical amplitude $\Omega_{\mathrm{c}}$ no steady state is formed such that a form of dynamical instability has arisen due to the driving overpowering the dissipation.
\\

\noindent \textit{Correlation functions}. \\
The impact of the $\mathrm{EP}$ is also felt in the $n$-th degree of coherence $g^{(n)}(\tau)$, of which we consider the first-order and second-order degrees in particular~\cite{Breuer2002, Gardiner2014}. The normalized first-order correlation function $g^{(1)}(\tau)$, with some delay time $\tau \ge 0$, may be defined as $g^{(1)}(\tau) = \lim_{t \to \infty} \langle b^\dagger (t) b (t+\tau) \rangle / \langle b^\dagger (t) b (t) \rangle $. For the presented model, using the results for the first and second moments and the quantum regression formula, it is explicitly given by either a damped-trigonometric expression or a damped-hyperbolic expression (see the Supplementary Information for details)
\begin{equation}
\label{geetwgjhgjgo3klj23}
 g^{(1)}(\tau) = \begin{cases}
  \left[ \cos \left(  \tilde{\omega} \tau \right) + \frac{\gamma}{2 \tilde{\omega}} \sin \left(  \tilde{\omega} \tau \right)  \right] \mathrm{e}^{- \frac{\gamma \tau}{2}},  & \quad\quad\quad\quad\quad \left( \Omega < \Omega_{\mathrm{EP}} \right),  \\
 \left[ \cosh \left(   \Gamma \tau \right) + \frac{\gamma}{2 \Gamma} \sinh \left(   \Gamma \tau \right)  \right] \mathrm{e}^{- \frac{\gamma \tau}{2} }, & \quad\quad\quad\quad\quad \left( \Omega > \Omega_{\mathrm{EP}} \right),
\end{cases}  
\end{equation}
where the frequency $\tilde{\omega}$ and the damping rate $\Gamma$ are defined in Eq.~\eqref{eq:Hamm213131232y} and Eq.~\eqref{eqapp:dfcbfcfbddsfdsfdfx} respectively, and where the characteristic time constant $\gamma/2$ appears in the decay factor. Clearly, in the regime of smaller driving amplitudes where $\Omega < \Omega_{\mathrm{EP}}$, the first-order coherence $g^{(1)}(\tau)$ is damped out at long timescales to zero,  $\lim_{\tau \to \infty } g^{(1)}(\tau) = 0$. However, in the opposing regime where $\Omega > \Omega_{\mathrm{EP}}$, the asymptotics are instead described by
\begin{equation}
\label{dfgfgdfb}
\lim_{\tau \to \infty } g^{(1)}(\tau) = \tfrac{1}{2} \left( 1 + \tfrac{\gamma}{2 \Gamma}  \right) \mathrm{e}^{- \left( \frac{\gamma}{2} - \Gamma \right) \tau },
 \quad\quad\quad\quad\quad\quad \left( \Omega > \Omega_{\mathrm{EP}} \right),
\end{equation}
which describes a convergent solution only when $\gamma / 2 > \Gamma$, or equivalently when the driving amplitude $\Omega < \Omega_{\mathrm{c}}$ [cf. Eq.~\eqref{eqapp:dfgdssdsfdfx}], which corresponds to when the system supports a steady state. Otherwise, a divergence in $g^{(1)}(\tau)$ with increasing time may be prevented by truncating the infinite energy ladder of the harmonic oscillator or by introducing anharmonicities. Notably, the formal definition of the normalized first-order correlation function $g^{(1)}(\tau)$, as given above, ensures that $g^{(1)}(0) = 1$ for all values of the driving amplitude $\Omega$.

\begin{figure*}[tb]
 \includegraphics[width=\linewidth]{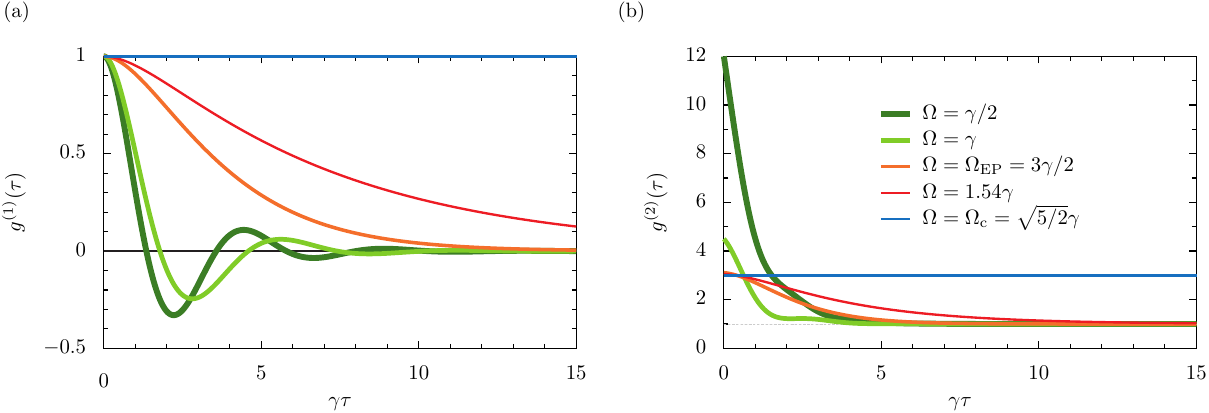}
 \caption{ \textbf{Correlations of the parametric driven-dissipative oscillator.} Panel (a): the degree of first-order coherence $g^{(1)}(\tau)$ as a function of the delay time $\tau$, in units of the inverse damping rate $\gamma^{-1}$ [cf. Eq.~\eqref{geetwgjhgjgo3klj23}]. We consider several values of driving amplitude $\Omega$, including below the exceptional point (green lines), exactly at the exceptional point (orange line) and above the exceptional point (red line), as described in the legend found in panel~(b). Panel (b): the degree of second-order coherence $g^{(2)}(\tau)$ as a function of $\tau$ [cf. Eq.~\eqref{sdfsfdsdf}]. Dashed grey line: $g^{(2)}(\tau) = 1$ for coherent light as a guide for the eye. In this figure, we consider the case of the detuning $\Delta = 3\gamma/2$, so that $\Omega_{\mathrm{EP}} = 3\gamma /2$ [cf. Eq.~\eqref{eqapp:fghn}] and $\Omega_{\mathrm{c}} = \sqrt{5/2}\gamma \simeq 1.58 \gamma$ [cf. Eq.~\eqref{eqapp:dfgdssdsfdfx}], a case which is given by the blue lines in both panels.}
 \label{geeone}
\end{figure*}

At the intermediate case between those described in Eq.~\eqref{geetwgjhgjgo3klj23}, that is when the $\mathrm{EP}$ is approached [cf. Eq.~\eqref{eqapp:fghn}], the first-order correlation function $g^{(1)}(\tau)$ collapses [in a similar manner to the population dynamics reconstruction of Eq.~\eqref{eqapp:adsads}] into the damped-quadratic form
\begin{equation}
\label{geetwgjhgjgodfdf3klj23}
 g_{\mathrm{EP}}^{(1)}(\tau) = \left( 1 + \tfrac{\gamma \tau}{2}  \right) \mathrm{e}^{- \frac{\gamma \tau}{2}},
 \quad\quad\quad\quad\quad\quad \left( \Omega = \Omega_{\mathrm{EP}} \right).
\end{equation}
We plot $g^{(1)}(\tau)$ as a function of the delay time $\tau$ in Fig.~\ref{geeone}~(a), which in general shows partial coherence with $0 < |g^{(1)}(\tau)| < 1$, as opposed to complete coherence with $|g^{(1)}(\tau)| = 1$ or full incoherence with $|g^{(1)}(\tau)| = 0$. We plot the results for driving amplitudes below the $\mathrm{EP}$ (green lines), at the $\mathrm{EP}$ (orange line), above the $\mathrm{EP}$ (red line) and at the critical driving amplitude (blue line). Notably, below the $\mathrm{EP}$ (two thickest lines) the correlation function displays characteristic damped oscillations in $\tilde{\omega} \tau$, and the exponential damping ensures $\lim_{\tau \to \infty} g^{(1)} (\tau) = 0$, and thus total incoherence, for all uncritical cases. Above the $\mathrm{EP}$ (red line) the behavior becomes, quite predictably, nonoscillatory. However, exactly at the critical driving amplitude $\Omega = \Omega_{\mathrm{c}}$, Eq.~\eqref{geetwgjhgjgo3klj23} reduces to $g_{\mathrm{c}}^{(1)}(\tau) = 1$ and complete coherence (thin blue line) is displayed for all times $\tau$.

Temporal light intensity correlations may be measured using the second order correlation function $g^{(2)}(\tau)$, which can likewise be defined in its normalized form as $g^{(2)}(\tau) = \lim_{t \to \infty} \langle b^\dagger (t) b^\dagger (t+\tau) b (t+\tau) b(t) \rangle / \langle b^\dagger (t) b (t) \rangle^2 $, for some delay time $\tau \ge 0$. Bunched light emissions arise from the oscillator when $g^{(2)}(0) > g^{(2)}(\tau)$, while photon antibunching occurs for $g^{(2)}(0) < g^{(2)}(\tau)$. The analytic expression for $g^{(2)}(\tau)$ may be calculated using the quantum regression formula and the first and second moments as (see the Supplementary Information for details)
 \begin{equation}
\label{sdfsfdsdf}
 g^{(2)} (\tau) = \begin{cases}
 1 +  \frac{\Delta^2 \left( 4 \tilde{\omega}^2 + \gamma^2 \right) + \Omega^2 \Big[ \left( 4 \tilde{\omega}^2 - \gamma^2 \right) \cos \left( 2 \tilde{\omega} \tau \right) + 4  \gamma \tilde{\omega} \sin \left( 2 \tilde{\omega} \tau \right) \Big]}{4 \tilde{\omega}^2 \Omega^2 } \mathrm{e}^{- \gamma \tau},  & \quad\quad\quad \left( \Omega < \Omega_{\mathrm{EP}} \right), \\
 1 +  \frac{\Delta^2 \left( 4 \Gamma^2 - \gamma^2 \right) + \Omega^2 \Big[ \left( 4 \Gamma^2 + \gamma^2 \right) \cosh \left( 2 \Gamma \tau \right) + 4  \gamma \Gamma \sinh \left( 2 \Gamma \tau \right) \Big]}{4 \Gamma^2 \Omega^2 } \mathrm{e}^{- \gamma \tau}, & \quad\quad\quad \left( \Omega > \Omega_{\mathrm{EP}} \right).
\end{cases}
\end{equation}
The expression valid for $\Omega < \Omega_{\mathrm{EP}}$ exhibits exponential decay with the time constant $\gamma$, such that $\lim_{\tau \to \infty} g^{(2)} (\tau) = 1$, suggesting a tendency towards Poissonian statistics for large delay times $\tau$. However, for the case of $\Omega > \Omega_{\mathrm{EP}}$ the long delay time asymptotics may be described by
\begin{equation}
\label{dsfsdfd}
 \lim_{\tau \to \infty} g^{(2)}(\tau) = 1 + \left( \tfrac{\Delta}{\Omega} \right)^2 \left[ 1 - \left( \tfrac{\gamma}{2\Gamma} \right)^2 \right] \mathrm{e}^{-  \gamma  \tau} + \tfrac{1}{2} \left( 1 + \tfrac{\gamma}{2\Gamma} \right)^2 \mathrm{e}^{- \left( \gamma - 2 \Gamma \right) \tau},  \quad\quad\quad \left( \Omega > \Omega_{\mathrm{EP}} \right),
\end{equation}
which describes a convergent solution only when $\gamma > 2 \Gamma$, or equivalently when $\Omega < \Omega_{\mathrm{c}}$ [cf. Eq.~\eqref{eqapp:dfgdssdsfdfx}], which corresponds to the situation when the system is able to support a steady state. At zero time delay ($\tau = 0$), we find from Eq.~\eqref{sdfsfdsdf} the much simpler expression
\begin{equation}
\label{geetwgjhgjdfdfdgodfdfddfdf3klj23}
 g^{(2)}(0) = 2 + \left( \frac{\Omega_{\mathrm{c}}}{\Omega} \right)^2,
\end{equation}
which has the upper bound $g^{(2)}(0) \simeq (\Omega_{\mathrm{c}}/ \Omega)^2$ for weak driving amplitudes $\Omega \ll \Omega_{\mathrm{c}}$, while the lower bound $g^{(2)}(0) = 3$ for strong drivings tending towards the critical amplitude $\Omega \to \Omega_{\mathrm{c}}$. These results effectively assume zero temperature of the thermal bath encasing the oscillator [cf. the discussion after the quantum master equation of Eq.~\eqref{eqapp:massdsdsdter}], however the order of limits is important in this case, and the expression of Eq.~\eqref{geetwgjhgjdfdfdgodfdfddfdf3klj23} should be replaced in the finite temperature case with a more complicated formula (as given in the Supplementary Information), such that one instead obtains the weak driving limit result $\lim_{\Omega \to 0} g^{(2)}(0) = 2$ for all temperatures, corresponding to photon bunching.

Exactly at the $\mathrm{EP}$, the following specific form of the correlation function $g^{(2)}(\tau)$ arises [cf. Eq.~\eqref{sdfsfdsdf} for the marginal cases directly below and above it]
\begin{equation}
\label{geetwgjhgjgodfdfddfdf3klj23}
 g_{\mathrm{EP}}^{(2)}(\tau) = 1 + \Big[ \left( \tfrac{\gamma}{2\Delta} \right)^2 + \tfrac{1}{2} \left( 2 + \gamma \tau\right)^2 \Big] \mathrm{e}^{- \gamma \tau},
 \quad\quad\quad\quad\quad \left( \Omega = \Omega_{\mathrm{EP}} \right),
\end{equation}
where again the $\mathrm{EP}$ has reconstructed the response of the system into a damped algebraic one. In Fig.~\ref{geeone}~(b), we plot $g^{(2)}(\tau)$ as a function of the delay time $\tau$, using the same colour coding as in panel (a). For drivings below the $\mathrm{EP}$ (green lines) we see gentle damped oscillations in the correlation function -- always satisfying the bunching inequality $g^{(2)}(0) > g^{(2)}(\tau)$. Exactly at the $\mathrm{EP}$ (orange line) we see the damped-quadratic scaling in $\gamma \tau$ following Eq.~\eqref{geetwgjhgjgodfdfddfdf3klj23} before a fast washing out of the correlations with large delay times $\tau \to \infty$, and a similar pattern occurs for cases above the $\mathrm{EP}$ (red line). Finally, at the critical driving amplitude $\Omega = \Omega_{\mathrm{c}}$ (blue line), where Eq.~\eqref{sdfsfdsdf} reduces to the critical result $g_{\mathrm{c}}^{(2)}(\tau) = 3$, the delay time-independent result is observed. All of these results have been obtained at zero temperature (see the Supplementary Information for a discussion of nonzero temperature).
\\

\begin{figure*}[tb]
 \includegraphics[width=\linewidth]{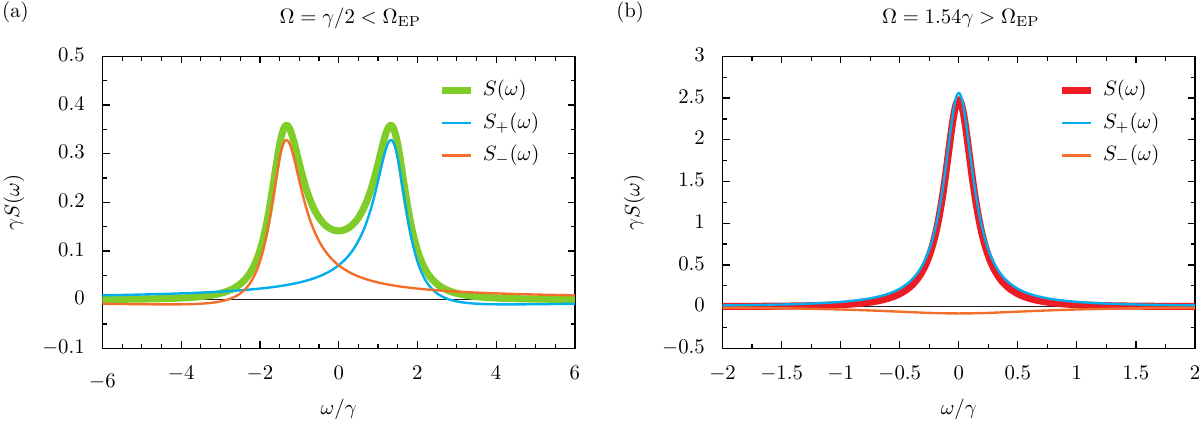}
 \caption{ \textbf{Optical spectrum of the parametric driven-dissipative oscillator.} The spectrum $S(\omega)$ (thick lines) of the emitted photons of frequency $\omega$ [cf. Eq.~\eqref{eqapp:dfgxcxcx}], and its decomposition into two constituent lineshapes $S_+(\omega)$ and $S_-(\omega)$ (thin cyan and orange lines respectively), all in units of the inverse loss rate $\gamma^{-1}$. Panel (a): the doublet regime, with the driving amplitude $\Omega = \gamma/2 < \Omega_{\mathrm{EP}}$. Panel (b): the singlet regime, with $\Omega = 1.54\gamma > \Omega_{\mathrm{EP}}$. In this figure, we consider the case of the detuning $\Delta = 3\gamma/2$, so that $\Omega_{\mathrm{EP}} = 3\gamma /2$ [cf. Eq.~\eqref{eqapp:fghn}] and $\Omega_{\mathrm{c}} = \sqrt{5/2}\gamma \simeq 1.58 \gamma$ [cf. Eq.~\eqref{eqapp:dfgdssdsfdfx}].}
 \label{spec}
\end{figure*}

\noindent \textit{Optical spectrum}. \\
The optical spectrum $S (\omega)$, describing the mean number of photons emitted from the oscillator with the frequency $\omega$, is defined via the integral of the population correlator $\langle b^\dagger (t) b (t+\tau) \rangle$, that is $S (\omega) = \lim_{t \to \infty} \pi^{-1} \langle b^\dagger (t) b (t) \rangle^{-1} \textrm{Re} \int_0^\infty \langle b^\dagger (t) b (t+\tau) \rangle \mathrm{e}^{\mathrm{i} \omega \tau} \mathrm{d}\tau $~\cite{Breuer2002, Gardiner2014, Downing20233}. We have included a normalization prefactor in this definition so that the spectral integral is equal to unity, that is $\int_{-\infty}^{\infty} S (\omega) \mathrm{d}\omega = 1$. We compute the analytic expression for the spectrum $S (\omega)$ of the parametric oscillator as (see the Supplementary Information for the derivation)
\begin{equation}
\label{eqapp:dfgxcxcx}
 S (\omega) = \begin{cases}
 \frac{\gamma}{\pi} \frac{\tilde{\omega}^2+\left( \frac{\gamma}{2} \right)^2}{ \Bigl\{ \left( \tfrac{\gamma}{2} \right)^2 + \left( \omega + \tilde{\omega} \right)^2 \Bigl\} \Bigl\{ \left( \tfrac{\gamma}{2} \right)^2 + \left( \omega - \tilde{\omega} \right)^2 \Bigl\} },  & \quad\quad\quad \left( \Omega < \Omega_{\mathrm{EP}} \right), \\
 \frac{\frac{1}{2}+\frac{\gamma}{4\Gamma}}{\pi} \frac{\tfrac{\gamma -  2 \Gamma}{2} }{ \left( \tfrac{\gamma - 2  \Gamma}{2} \right)^2 + \omega^2  } +  \frac{\frac{1}{2}-\frac{\gamma}{4\Gamma}}{\pi} \frac{\tfrac{\gamma +  2 \Gamma}{2} }{ \left( \tfrac{\gamma + 2  \Gamma}{2} \right)^2 + \omega^2  }, & \quad\quad\quad \left( \Omega > \Omega_{\mathrm{EP}} \right),
\end{cases}
\end{equation}
where the key quantities $\tilde{\omega}$ and $\Gamma$ were introduced in Eq.~\eqref{eqapp:umadfdfsdsdsdsdstrix} and Eq.~\eqref{eqapp:dfcbfcfbddsfdsfdfx} respectively. This spectral result $S (\omega)$ is plotted in Fig.~\ref{spec} as the thick lines. In panel (a), the driving amplitude $\Omega$ is below the $\mathrm{EP}$ ($\Omega < \Omega_{\mathrm{EP}}$), such that the optical spectrum presents a characteristic doublet lineshape (thick green line). This response may have been anticipated due to the presence of two distinct real parts of the complex eigenfrequencies $\omega_{\pm}$ in this regime [cf. Fig.~\ref{excep}~(b)]. However, for the case of Fig.~\ref{spec}~(b), where the driving amplitude is above the $\mathrm{EP}$ ($\Omega > \Omega_{\mathrm{EP}}$), the spectrum displays a singlet structure (thick red line) since there is now only one distinct real part of $\omega_{\pm}$, which is essentially the only transition frequency appearing in the system [cf. Fig.~\ref{excep}~(b)]. In both panels of Fig.~\ref{spec}, alongside the full spectrum $S (\omega)$ we plot its decomposition into two parts $S_+ (\omega)$ (thin cyan lines) and $S_- (\omega)$ (thin orange lines), where each contribution $S_\pm (\omega)$ corresponds to an allowed transition in the system (see the Supplementary Information for further information). In panel (a) we therefore see one peak centered around $\tilde{\omega}$ and the other peak at $-\tilde{\omega}$, while in panel (b) both peaks are at resonance. The presence of a singlet lineshape within the lower expression within Eq.~\eqref{eqapp:dfgxcxcx} can be more easily seen exactly at the $\mathrm{EP}$, where the optical spectrum $S (\omega)$ reduces to the more compact result
\begin{equation}
\label{eqapp:dfgxcdfdfxcx}
 S_{\mathrm{EP}} (\omega) = \frac{2}{\pi} \frac{ \left( \frac{\gamma}{2} \right)^3}{ \left[ \left( \frac{\gamma}{2} \right)^2 + \omega^2 \right]^2 },
  \quad\quad\quad\quad\quad\quad \left( \Omega = \Omega_{\mathrm{EP}} \right).
\end{equation}
This expression is of the form of a probability density function for a certain type of Student's $t$-distribution with the degree of freedom $\nu = 3$ (notably, a Lorentzian distribution corresponds to $\nu = 2$, and a Gaussian distribution arises in the limit of $\nu \to \infty$)~\cite{Abramowitz1972}. Hence the optical spectrum $S (\omega)$ presents one of the most tangible indicators of passing through an $\mathrm{EP}$, due the stark difference in spectral features presented, namely a doublet to singlet transition.
\\

\begin{figure*}[tb]
 \includegraphics[width=\linewidth]{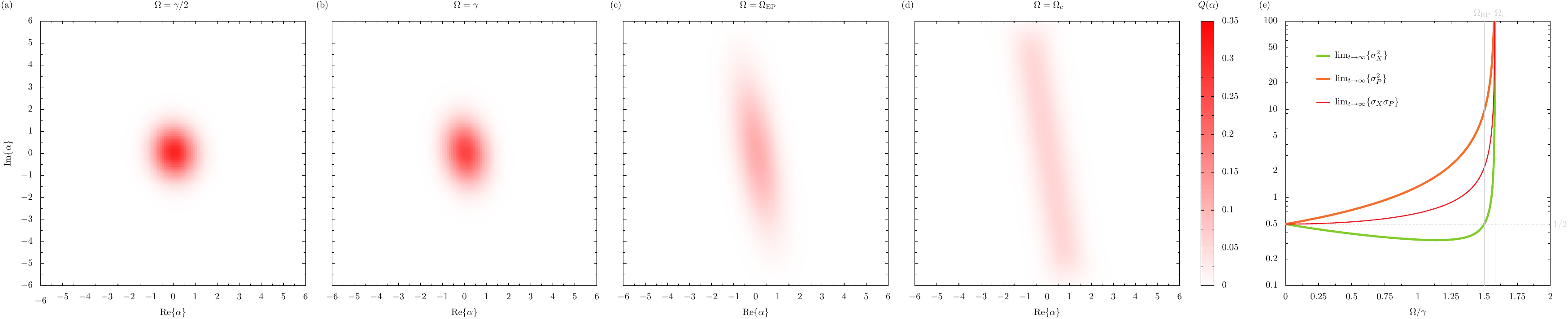}
 \caption{ \textbf{The phase-space quasiprobability distribution and variances of the parametric driven-dissipative oscillator.} Panels (a--d): the Husimi function $Q \left( \alpha \right) = \bra{\alpha} \hat{\rho} \ket{\alpha}/\pi$, or the expectation value of the density operator $\hat{\rho}$ with respect to the coherent state $\ket{\alpha}$, for several values of the driving amplitude $\Omega$. The $Q$-function is calculated in the steady state ($t \to \infty$) for a truncated oscillator with $N = 40$ levels. Panel (a): $\Omega = \gamma/2$, where $\gamma$ is the loss rate. Panel (b): $\Omega = \gamma$. Panel (c): $\Omega = 3\gamma/2$, corresponding to the exceptional point $\Omega_{\rm{EP}}$ [cf. Eq.~\eqref{eqapp:fghn}]. Panel (d): $\Omega = \sqrt{5/2}\gamma \simeq 1.58 \gamma$, corresponding to the critical point $\Omega_{\rm{c}}$ [cf. Eq.~\eqref{eqapp:dfgdssdsfdfx}]. Panel (e): the quadrature variances $\sigma_X^2$ (thick green line) and $\sigma_P^2$ (thick orange line) in the steady state, as a function of $\Omega$. The product of the standard deviations $\sigma_X \sigma_P$ (medium red line) is also shown, as is a guide for the eye at the Robertson-Schr\"{o}dinger minimum uncertainty of $1/2$ (horizontal, dashed grey line). Vertical grey lines: driving amplitudes corresponding to $\Omega_{\rm{EP}}$ and $\Omega_{\rm{c}}$ respectively. Throughout this figure, we consider the case of the detuning $\Delta = 3\gamma/2$.}
 \label{qfunc40}
\end{figure*}

\noindent \textit{Quantum states}. \\
The quantum states of the system can be analysed within a phase-space formalism~\cite{Cahill1969, Hillery1984}. In particular, the expectation value of the density operator $\hat{\rho}$, with respect to the coherent state $\ket{\alpha}$, is the so-called Husimi function $Q \left( \alpha \right) = \bra{\alpha} \hat{\rho} \ket{\alpha}/\pi$. This $Q$-function acts like a kind of phase-space quasiprobability distribution, since it has the normalization $\int Q \left( \alpha \right) \mathrm{d}^2 \alpha = 1$ and it is non-negative for all quantum states. In Fig.~\ref{qfunc40} we show the evolution of the $Q$-function with increasing driving amplitude $\Omega$ across the row of panels (in the steady state $t \to \infty$, where the detuning $\Delta = 3\gamma/2$ and for a truncated oscillator with $N = 40$ levels). In panel (a) the driving amplitude $\Omega = \gamma/2$, which is sufficiently small such that the $Q$-function is approximately circular since the quantum state is essentially brought into the vacuum state with $n = 0$ due to the dissipation (approximately, here $Q \left( \alpha \right) \simeq \mathrm{e}^{-|\alpha|^2} /\pi$ and its maximum $\mathrm{max} \{ Q \} = 1/\pi \simeq 0.318$). With increasing drive amplitude up to $\Omega = \gamma$ in panel (b), the $Q$-function becomes increasingly oval-shaped in the first manifestations of the parametric driving inducing squeezing behavior. In panel (c) the situation at the $\mathrm{EP}$ is shown ($\Omega = \Omega_{\rm{EP}} = 3\gamma/2$), and the significance of the squeezing Hamiltonian of Eq.~\eqref{eq:Hammy} is clearly seen through the highly squeezed $Q$-function (the tilt is caused by the competition between the driving and the dissipation). Finally in panel (d), where the critical point is finally reached ($\Omega = \Omega_{\rm{c}} = \sqrt{5/2}\gamma \simeq 1.58 \gamma$), the $Q$-function becomes stadium-shaped (and rather diluted) due to the finite truncation (see the Supplementary Information for more details). While the results of Fig.~\ref{qfunc40}~(a-c) are essentially those for an oscillator with an infinite number of levels, the truncation to $N = 40$ levels is necessary in panel (d) since the driving amplitude there coincides with the critical point in the untruncated limit [where no steady state exists, as suggested by Eq.~\eqref{eqapp:dfgdsfdfx}].

The quasiprobability distribution results of Fig.~\ref{qfunc40}~(a--d) can be better understood in conjunction with an analysis of the variances $\sigma_X^2$ and $\sigma_P^2$ of the generalized quadrature operators $\hat{X}$ and $\hat{P}$ [cf. the discussion above Eq.~\eqref{eqapp:massdsdsdter}]. The two quadrature variances $\sigma_X^2 = \langle \hat{X}^2 \rangle - \langle \hat{X} \rangle^2$ and $\sigma_P^2 = \langle \hat{P}^2 \rangle - \langle \hat{P} \rangle^2$ should then satisfy, in their standard deviation forms $\sigma_X$ and $\sigma_P$, the Robertson-Schr\"{o}dinger uncertainty relation $\sigma_X \sigma_P \ge 1/2$. In the steady state ($t \to \infty$), we find the following simple expressions for the variances (see the Supplementary Information for the derivation)
\begin{equation}
\label{eq:sfdsdfsfsfsfd}
\lim_{t \to \infty} \sigma_{X}^2 = \frac{1}{2} \frac{ \Omega_{\mathrm{c}}^2 - \Omega \Delta }{ \Omega_{\mathrm{c}}^2 - \Omega^2 },
\quad\quad\quad\quad\quad\quad\quad\quad
\lim_{t \to \infty} \sigma_{P}^2 = \frac{1}{2} \frac{ \Omega_{\mathrm{c}}^2 + \Omega \Delta }{ \Omega_{\mathrm{c}}^2 - \Omega^2 },
\end{equation}
where the critical driving amplitude $\Omega_{\mathrm{c}}$ is given by Eq.~\eqref{eqapp:dfgdssdsfdfx}, and where Eq.~\eqref{eq:sfdsdfsfsfsfd} is defined for driving amplitudes $\Omega < \Omega_{\mathrm{c}}$ so that a steady state limit exists. We plot the variances $\sigma_{X}^2$ (thick green line) and $\sigma_{P}^2$ (thick orange line), as a function of $\Omega$ in Fig.~\ref{qfunc40}~(e). Most notably, for drivings $\Omega < \Omega_{\mathrm{EP}}$ the position variance $\sigma_{X}^2 < 1/2$, showcasing the same steady-state squeezing as displayed pictorially in Fig.~\ref{qfunc40}~(a--c). In particular, Eq.~\eqref{eq:sfdsdfsfsfsfd} implies that the position variance minimum, $\mathrm{min} \{ \sigma_{X}^2 \} = [ 1 + \gamma/(2\Omega_{\mathrm{c}}) ]/4$, occurs when $\Omega = \Omega_{\mathrm{c}} ( \Omega_{\mathrm{c}} - \gamma/2 ) / \Delta$. For the case of Fig.~\ref{qfunc40}~(e), where the detuning parameter $\Delta = 3\gamma/2$, this means that the extremum $\mathrm{min} \{ \sigma_{X}^2 \} =  (1+1/\sqrt{10})/4 \simeq 0.329$ is realized when the driving amplitude $\Omega = \gamma (10-\sqrt{10})/6  \simeq 1.14 \gamma$, which is the case somewhere between those described in panels (b) and (c). Notably, exactly at $\Omega = \Omega_\mathrm{EP}$ the position variance $\sigma_{X}^2 = 1/2$, suggesting the role of the $\rm{EP}$ as the threshold above which steady state squeezing is absent (this coincidence is seemingly accidental and does not hold at sufficiently high temperatures of the thermal bath, as is shown in the Supplementary Information). Finally, the product of the standard deviations $\sigma_X \sigma_P$ (medium red line) has the lower bound $1/2$ in the vanishing driving limit $\Omega \to 0$ and tends to infinity in the maximum driving limit of $\Omega \to \Omega_{\mathrm{c}}$, marking the onset of the dynamical instability associated with the loss of the steady state.
\\

\begin{figure*}[tb]
 \includegraphics[width=\linewidth]{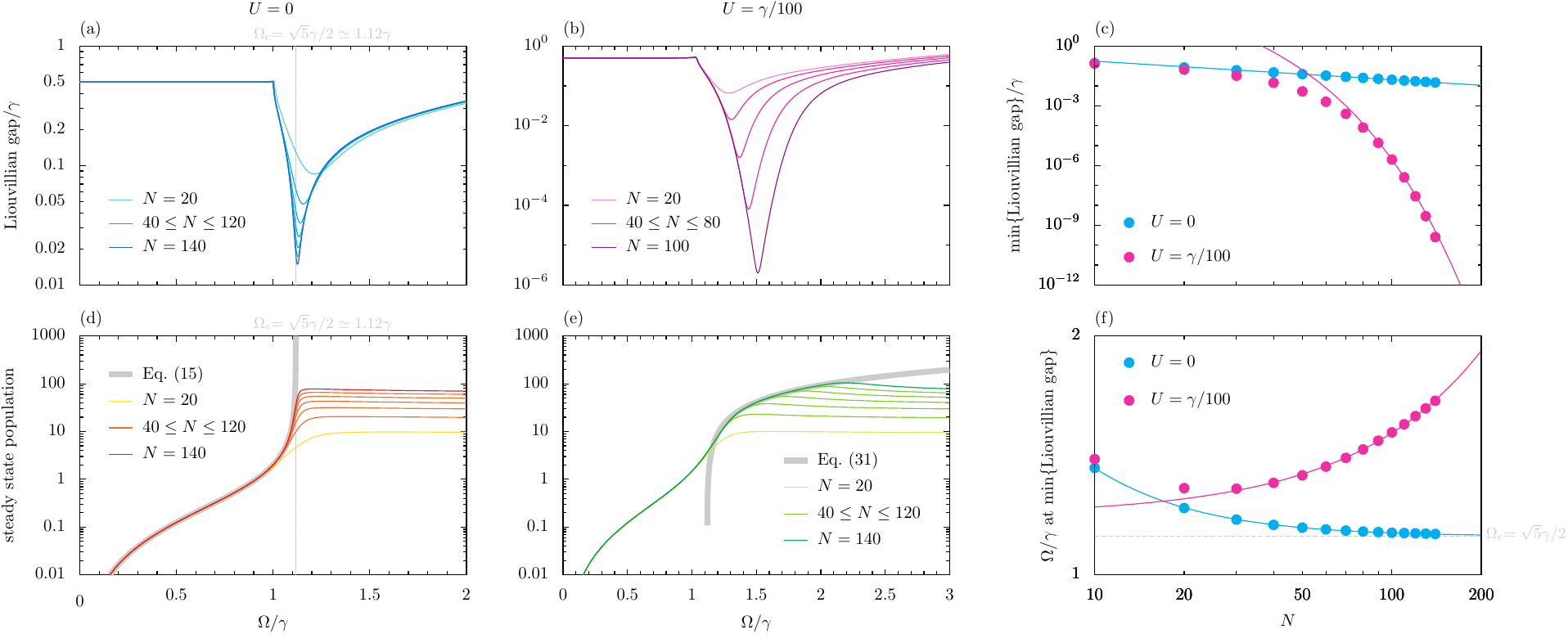}
 \caption{ \textbf{The effect of interactions on the parametric driven-dissipative oscillator.} Panel (a): the Liouvillian gap as a function of the drive amplitude $\Omega$ (in units of the loss rate $\gamma$) for the harmonic case ($U = 0$). Results are shown for various cases when the oscillator is truncated to have a finite number of levels $N$. Vertical line: the gap closes as $\Omega \to \Omega_{\mathrm{c}}$ with $N \to \infty$ [cf. Eq.~\eqref{eqapp:dfgdssdsfdfx}]. Panel (b): as for panel (a), but for a typical anharmonic case ($U = \gamma/100$). Panel (c): the minimum value of the Liouvillian gap as a function of the system size $N$. The algebraic scaling for the harmonic case (cyan line) is $\gamma A/N^B$, with $\{ A, B \} = \{ 1.505, 0.933 \}$. The anharmonic case (pink line) has the exponential scaling $\gamma A/B^N$, with $\{ A, B \} = \{ 1941, 1.230 \}$. Panel (d): the mean steady state population of the oscillator $\lim_{t \to \infty} \langle b^\dagger b \rangle$ as a function of $\Omega$, computed for the harmonic case ($U = 0$). Thick grey line: the untruncated oscillator expression of Eq.~\eqref{eqapp:dfgdsfdfx}. Panel (e): as for panel (d), but for the considered anharmonic case ($U = \gamma/100$). Thick grey line: the semiclassical expression of Eq.~\eqref{geetwgdfdfsdsdfddfdfdfdfdfdfjhgdfjgo3klj23}. Panel (f): the value of $\Omega/\gamma$ at which $\textrm{min \{Liouvillian gap\} }$ occurs as a function of $N$. The harmonic case (cyan line) approaches $\Omega_{\mathrm{c}}$ via the algebraic scaling $\Omega = \Omega_{\mathrm{c}} + \gamma A /N^B$, with $\{ A, B \} = \{ 5.282, 1.333 \}$. The anharmonic case (pink line) diverges exponentially like $\Omega = \gamma A B^N$, with $\{ A, B \} = \{ 1.189, 1.0024 \}$. In this figure, we consider the case of the detuning $\Delta = \gamma$, so that $\Omega_{\mathrm{c}} = \sqrt{5}\gamma /2 \simeq 1.12 \gamma$ [cf. Eq.~\eqref{eqapp:dfgdssdsfdfx}] for the harmonic case.}
 \label{horn}
\end{figure*}

\noindent \textit{Liouvillian gap}. \\
The Liouvillian eigenmatrix $\mathcal{L}$, corresponding to the model of Eq.~\eqref{eqapp:massdsdsdter} truncated into its finite-dimensional matrix form $\partial_t \rho = \mathcal{L} \rho$, possesses a certain number of eigenvalues (see the Supplementary Information for more details). From this, one may obtain the Liouvillian spectral gap (formally, the Liouvillian eigenvalue with the smallest real part after discounting any zero eigenvalues)~\cite{Kessler2012, Cai2013, Minganti2018}. A closing of the Liouvillian gap at some critical value of a system parameter signifies a dissipative phase transition, examples of which have recently been observed in several photonic platforms~\cite{Fitzpatrick2017, Rodriguez2017, Fink2018}. We plot the Liouvillian gap as a function of the drive amplitude $\Omega$ for the studied squeezed oscillator in Fig.~\ref{horn}~(a). We show results for an increasingly large number of levels $N$ of the oscillator (thin cyan-blue lines), such that in the untruncated limit -- that is, the infinite limit $N \to \infty$ as is written in Eq.~\eqref{eq:Hammy} -- the Liouvillian gap closes. This closing of the gap occurs at the critical amplitude $\Omega_{\mathrm{c}}$ (vertical grey line), as was previously defined in Eq.~\eqref{eqapp:dfgdssdsfdfx}, and which signifies a dissipative phase transition. Furthermore, the purity $\mathcal{P} = \Tr (\rho)^2$ of the density matrix $\rho$ in the steady state suggests that while below $\Omega_{\mathrm{c}}$ the state is relatively pure ($\mathcal{P} \simeq 1$), sharply above $\Omega_{\mathrm{c}}$ the state is maximally mixed ($\mathcal{P} \simeq 1/N$), as is shown in the Supplementary Information. In the panel directly below, Fig.~\ref{horn}~(d), we plot the average population of the oscillator in the steady state, $\lim_{t \to \infty} \langle b^\dagger b \rangle$, again as a function of $\Omega$ and for different level numbers $N$ (thin yellow-red lines) due to the truncation. These restricted results approach the behaviour of the infinite-level oscillator result [thick grey line, Eq.~\eqref{eqapp:dfgdsfdfx}] with higher $N$, revealing that the critical amplitude $\Omega_{\mathrm{c}}$ is indeed linked to the disappearance of a steady state population (or at least a saturation for finite $N$ cases) as well as a dissipative phase transition.
\\

\noindent \textit{A note on interactions}. \\
Anharmonic deviations to the harmonic oscillator described in Eq.~\eqref{eq:Hammy} can be modelled by adding a Kerr-like energy to the Hamiltonian, $\hat{H} \to  \hat{H} + \hat{H}_{\mathrm{I}}$, with the help of the nonlinear term
\begin{equation}
\label{eq:Hammddfdfy}
 \hat{H}_{\mathrm{I}} = \frac{U}{2} b^{\dagger} b^{\dagger} b b,
\end{equation}
where the on-site interaction strength $U \ge 0$ is considered to be repulsive. Carrying out an analysis of the Liouvillian gap of this anharmonic model, again for a spread of different truncations to $N$ energy levels (pink-purple lines), leads to the results of Fig.~\ref{horn}~(b). Notably, unlike the harmonic case of panel~(a), the Liouvillian gap in panel (b) seems to decrease with increasing $N$ without a convergence to some critical value of the driving amplitude $\Omega$ in the large $N$ thermodynamic limit. This statement is supported by Fig.~\ref{horn}~(c, f). In panel (c), where the minimum of the Liouvillian gap is shown as a function of the restriction parameter $N$, for the anharmonic case (pink circles) and harmonic case (cyan circles), we see how the Liouvillian gap is closing as a power function and exponentially respectively (coloured lines) via the presented fittings. Meanwhile in panel (f), which displays the value of the drive amplitude $\Omega$ at which $\textrm{min \{Liouvillian gap\} }$ occurs as a function of $N$, the fittings suggest that the harmonic case (cyan line) approaches $\Omega_{\mathrm{c}}$ via a power function, while the anharmonic case (pink line) seems to diverge exponentially. This apparent absence of a dissipative phase transition at some critical system parameter for the anharmonic oscillator comes with the guarantee of a steady state population in this case. Indeed, a semiclassical analysis~\cite{Meaney2014} of this nonlinear model [cf. Eq.~\eqref{eq:Hammy} with Eq.~\eqref{eq:Hammddfdfy}] leads to the following expression for the steady state population $n_{\mathrm{I}} \left( \infty \right)$ of the anharmonic oscillator (see the Supplementary Information for the derivation)
\begin{equation}
\label{geetwgdfdfsdsdfddfdfdfdfdfdfjhgdfjgo3klj23}
n_{\mathrm{I}} \left( \infty \right) \simeq  \frac{1}{U} \bigg(  \sqrt{ \Delta^2 + \Omega^2 - \Omega_{\mathrm{c}}^2    } - \Delta \bigg),
   \end{equation}
 where the critical frequency $\Omega_{\mathrm{c}}$ is defined in Eq.~\eqref{eqapp:dfgdssdsfdfx} (cf. Eq.~\eqref{eqapp:dfgdsfdfx} for the corresponding harmonic oscillator population result). Clearly, it is necessary (within this semiclassical analysis) for the driving amplitude to be sufficiently strong $\Omega >\Omega_{\mathrm{c}}$, in order to ensure a physical steady state with $n \left( \infty \right) > 0$. The population of Eq.~\eqref{geetwgdfdfsdsdfddfdfdfdfdfdfjhgdfjgo3klj23} is plotted in Fig.~\ref{horn}~(e) as the thick grey line, along with results for various truncations $N$ of the anharmonic oscillator (thin green lines). This panel suggests that Eq.~\eqref{geetwgdfdfsdsdfddfdfdfdfdfdfjhgdfjgo3klj23} is indeed a reasonable approximation in the thermodynamic limit $N \to \infty$ of a large number of excitations. We conclude that while interactions in the form of Eq.~\eqref{eq:Hammddfdfy}, at least for perturbative values of $U$, do not strongly modify the majority of our results, this anharmonicity is important for the extinction of the critical closing of the Liouvillian gap and hence for the absence of a dissipative phase transition. 
\\


\noindent \textbf{Discussion}\\
In conclusion, we have studied one of the simplest driven-dissipative quantum models in which exceptional points can arise -- that of a quantum harmonic oscillator with parametric driving. We have revealed that the exceptional point is of second-order in the first moments of the system, which impacts upon the lineshape of the optical spectrum and the character of the first-order degree of coherence. In the second moments of the system the exceptional point instead appears at the third-order, which influences both the dynamics of the populations and the behaviour of the second-order degree of coherence. Furthermore, the exceptional point is shown to coincide with last remnants of quantum squeezing in the steady state, which perhaps highlights the importance of this kind of physics for quantum states, as described within an open quantum systems approach. We have also discussed the occurrence of a critical point for the parametric oscillator which is associated with a dynamical instability, the phenomenon of a dissipative phase transition (as adjudicated by the closing of the Liouvillian gap), and the impact of small anharmonicities and truncations of the oscillator. Overall, we hope that our results can stimulate further experimental work, with platforms including photonic cavities~\cite{Rota2019, Marty2021} and quantum circuits~\cite{Macklin2015, Nigg2017}, which can observe the predicted non-Hermitian quantum physics through the insightful lens of exceptional points and critical points. 
\\


\noindent \textbf{Methods}\\
We use methods from theoretical quantum optics as outlined in the main text and as discussed in detail in the Supplementary Information.
\\


\noindent \textbf{Acknowledgments}\\
\textit{Funding}: CAD is supported by the Royal Society via a University Research Fellowship (URF\slash R1\slash 201158) and a Royal Society Research Grant (RGS\slash R1 \slash 211220). AVB is supported by the CNPq (Conselho Nacional para o Desenvolvimento Científico e Tecnológico) with grant number 465469/2014-0. \textit{Data and materials availability}: All data is available in the manuscript and the Supplementary Information.
\\


\noindent \textbf{Author contributions}\\
CAD conceived of the study, performed the calculations and wrote the first version of the manuscript. AVB suggested carrying out the phase-space and quadrature analyses, held fruitful discussions with CAD and helped to revise the manuscript. Both authors gave final approval for publication.
\\


\noindent
\textbf{ORCID}\\
C. A. Downing: \href{https://orcid.org/0000-0002-0058-9746}{0000-0002-0058-9746}.
\\
A. Vidiella-Barranco: \href{https://orcid.org/0000-0002-6918-8764}{0000-0002-6918-8764}.
\\


\noindent \textbf{Data availability}\\
There is no additional data. Further information is given in the Supplementary Information.
\\


\noindent \textbf{Competing interests}\\
The authors declare no competing interests.
\\



\end{document}